\documentclass[graphicx, 12pt,preprint,groupedaddress]{revtex4-1}
\usepackage{textpos}  
\usepackage{hyperref}
\hypersetup{colorlinks=true, citecolor=blue, urlcolor=blue, linkcolor=blue}
\usepackage{graphicx}
\begin{document}
\begin{textblock*}{\textwidth}(0cm,0cm)
{This article may be downloaded for personal use only. Any other use requires prior permission of AIP Publishing. {\it This is the accepted author version of the article} that appeared in Applied Physics Letters vol. 74, 3933 (1999) [$\copyright$ 1999 American Institute of Physics]. The version of record is available from the publisher's website from the link: \url{https://aip.scitation.org/doi/10.1063/1.124228}}  
\end{textblock*}  
\vspace{4cm}  
\title{\bf Optical Photonic Crystals fabricated from colloidal systems }
\author{G. Subramania}
\address{Ames Laboratory-USDOE and Department of Physics and Astronomy 
Iowa State University, Ames, Iowa 50011} 
\author{K. Constant} 
\address{Ames Laboratory-USDOE and 
Department of Materials Science and Engineering,
Iowa State University, Ames, Iowa 50011} 
\author{R. Biswas}
\author{M. M. Sigalas}
\author{K.-M. Ho}
\address{Ames Laboratory-USDOE, Microelectronics Research Center, and
Department of Physics and Astronomy, Iowa State University, Ames, Iowa 50011} 

\vspace{2cm}
\begin{abstract}  
Photonic crystals of close-packed arrays of air spheres in a dielectric background
of titania  have been fabricated with a novel ceramic technique.
Unlike previous methods, ordering of the spheres and the formation of
the titania network are performed simultaneously. The photonic crystals
exhibit a reflectance peak and a uniform color
at the position of the first stop band. The
wavelength of the reflectance peak scales very well with the sphere size.
\end{abstract}

\maketitle

Photonic crystals (PC) are structures in which
the dielectric constant varies periodically in space.
If the contrast in dielectric constant is large enough,
a frequency gap may occur in which light cannot propagate
analogous to the electronic band gap in semiconductors.
\cite{Yablonovitch, John, Soukoulis}
The emission of light inside a photonic crystal
can be strongly manipulated
in the region of the photonic band gap, resulting in
intense interest in optoelectronics and photocatalysis applications.
However, the fabrication of such crystals at visible
wavelengths has posed a formidable challenge. Recently,
several groups \cite{Vos,Imhof,Velev,Zakhidov,Holland}
have successfully fabricated 
materials in which pores are arranged in periodic arrays on 
a length scale comparable to optical wavelengths. Such
materials are very good candidates for use as photonic
crystals. However, in all the previous publications, a 
definitive signature of the existence of a photonic gap
is still missing.
We present a novel ceramic fabrication technique 
for photonic crystal thin films at visible wavelengths.
We show that the reflectance spectra of these crystals shift 
systematically with the pore size, providing
evidence of photonic crystal effects.

Colloidal crystals are very attractive candidates for 
use in fabrication of optical photonic crystals.
Monodisperse colloidal suspensions of silica or polystyrene spheres
can self-assemble into close-packed structures at optical length scales,
with excellent long-range periodicity.
However, to observe photonic gaps requires an inverse structure where 
lower dielectric spheres (refractive index $n_1$) are embedded in an 
interconnected higher dielectric background (refractive index $n_2$).
Optimum photonic effects require a low filling ratio (20-30 $\%$) of the 
dielectric background.\cite{theory}
A major difficulty lies in the introduction of an interconnected
dielectric background for the colloidal spheres, and the subsequent 
removal of the spheres,
by calcination or etching, to achieve the desired dielectric contrasts.

We use titania as the background dielectric filling medium which has a refractive index of 
$\sim 2.6 -2.8$ at optical wavelengths, with negligible absorption above 400 nm.
In contrast to previous work \cite{Vos, Zakhidov, Holland} 
where the colloidal template is first assembled and the 
titania is introduced afterwards in a sol-gel process, we start with a 
slurry of nano-crystalline titania suspension and monodisperse polystyrene spheres. 
A few drops of this slurry is spread on a glass substrate and allowed to dry slowly over 
a period of $\sim 24$ hours in a humidity chamber. 
The samples are then pressed in a cold isostatic press to
improve the initial green density of the as-dried samples and to reduce stress cracks
during subsequent heat treatment. 
The sample is slowly heated to $520^oC$ for $5$ hours whereby the polystyrene 
spheres are burned off leaving behind air spheres in a titania matrix.
Thin films with dimensions $\sim 10mm\times 2-3mm$  can be reproducibly synthesized in this way in much shorter times ($\sim$1 day) 
than with the infiltration technique.\cite{Vos, Zakhidov, Holland}

Optical inspection of our samples reveals shiny regions with characteristic
colors that depends on the size of the polystyrene spheres used. 
This is especially clear when the samples are viewed under the microscope.
Samples fabricated with 395 nm spheres exhibit bright green regions.
With larger spheres (479 nm), the color shifts
to a salmon-red color. 
Unlike previous reports\cite{Vos, Zakhidov, Holland}
our films exhibit uniform color over large regions millimeters in size.

Wide view scanning electron microscope (SEM) images (Fig. 1a) show large domains 
with excellent order extending from $\sim 50 \mu m$ to more than $\sim 100 \mu m$.
Also visible (Fig. 1a) are single-height steps separating large domains of hexagonally
ordered
regions. The domains are well ordered across drying cracks in the
sample (Fig. 1a), indicating that ordering in the samples occurs upon deposition and
is not disrupted by the drying and heating process.
Our crystals exhibit considerably better short-range and long-range order 
than the macroporous materials fabricated with sol-gel methods\cite{Imhof}.

At still lower magnification, the scan periodicity of the CRT display and the object periodicity
interact, producing fringes or a Moire pattern.\cite{Goldstein}
This pattern (Fig. 1b) illustrates well the domain orientation and strain within the
individual domains.

Higher magnification SEM images (Fig. 1c) reveals  
hollow regions of air spheres that are very well ordered in a triangular
lattice. 
There are three dark regions inside each 
hollow region corresponding to the air spheres of the underlying layer, indicating that the
spheres are indeed close-packed. 
The SEM images indicate that the crystalline grains in
the film are highly oriented with the close-packed planes parallel to
the substrate. Preferential orientation also exists in the close-packed
plane probably due to stresses developing during the
drying process. This alignment of crystal grains may prove very useful
for applications and measurements especially in cases where a full
photonic band gap does not exist in all directions of propagation in
the crystal.
Determination of the lattice constant indicates a small shrinkage of
$\leq 5 \%$ in the lateral direction of the film due to the heat treatment
and densification of the titania network. 
Experimental thickness measurements, before and after the pressing and heating
process, indicate a larger shrinkage in the direction perpendicular to the film.



Since the ordered films are thick ($\geq 10 \mu$), their transmission is small and the major
optical signature is found in reflectance measurements.
The specular reflectance at near normal incidence 
from our nanostructured films
is shown (Fig. 2a)  for different sizes of polystyrene spheres as templates. 
The initial sphere sizes (Fig. 2a, legend) were measured directly from SEM images 
of ordered arrays of polystyrene spheres.

The prominent feature is a specular reflectivity peak for each structure, that 
systematically shifts from 1120 nm to 521 nm over the range of photonic crystals.
The wavelength of the specular peak corresponds very well to the visual color of the
samples. 
The larger pore samples have reflectivity peaks in the near-infrared. 
In addition, there is a gradual but featureless
increase in reflectivity at longer wavelengths (above 1000 nm) in several samples
(Fig. 3a). This is due to the rough surface of the PC's appearing
smoother when probed at longer wavelengths. This increases the specular
reflectivity, with an accompanying decrease of the diffuse reflectance at longer
wavelength, which is also observed.

The position of the observed reflectivity peak
scales remarkably well with the diameter of the spheres (Fig. 2b), 
indicating that it 
is an intrinsic feature of the photonic crystals. 
This is the first observation of the optical signature of a photonic crystal
together with the required scaling with sphere size, 
that has not been seen in any previous work \cite{Vos,Imhof,Velev,Zakhidov,Holland}
on such templated PC's. 

We performed
photonic band calculations and calculated reflectivities from transfer
matrix simulations\cite{later}, and find that the peak arises 
from the wide stop band in the stacking direction 
for close-packed structures. 
For the fcc structure this corresponds to the stop band between the lowest
bands 2 and 3.
Our calculations find that the existence and position of the stop band in 
the stacking direction
are insensitive to the stacking sequence of the spheres 
(fcc (ABC) or hcp (ABAB)) \cite{later}.
The stop band corresponds to the known pseudogap in the photonic densities of 
states, and
persists even for lower n $\sim$ 2 over a large range of filling fractions.
Such a refractive index for titania may be expected from the considerably
lower density of the solid titania matrix, as would be expected from these processing conditions and from earlier sintering studies
of nanocrystalline titania\cite{Hahn}.
Further porosimetry measurements \cite{later}, will be performed to 
estimate the density of the titania matrix.
The refractive index of the background skeletal titania may be improved by
sintering at higher temperatures\cite{Hahn}. 
Quantitative calculations of peak wavelengths will be presented later, when
accurate interlayer spacings are determined from optical diffraction 
measurements, but preliminary estimates indicate that the observed peak
frequencies are consistent with what we know about the geometry and filling
fraction of the films.

  The success in the fabrication of large-area
optical photonic crystals using rapid, economical, and reproducible ceramic 
techniques will open the way towards the experimental observation of many
interesting effects involving the control of light emission and propagation
in these materials.

We thank J. Kavanaugh, G. Tuttle, P. Canfield, A. Panchula,
H. Kang, and W. Leung for help with various measurements and C. M. Soukoulis
and S. John for helpful discussions.
This research was supported by the Office
of Basic Energy Sciences, and Ames Laboratory is operated for the U.S. DOE by
Iowa State University.

\newpage

\begin{figure}
\centering
\includegraphics[width=8cm]{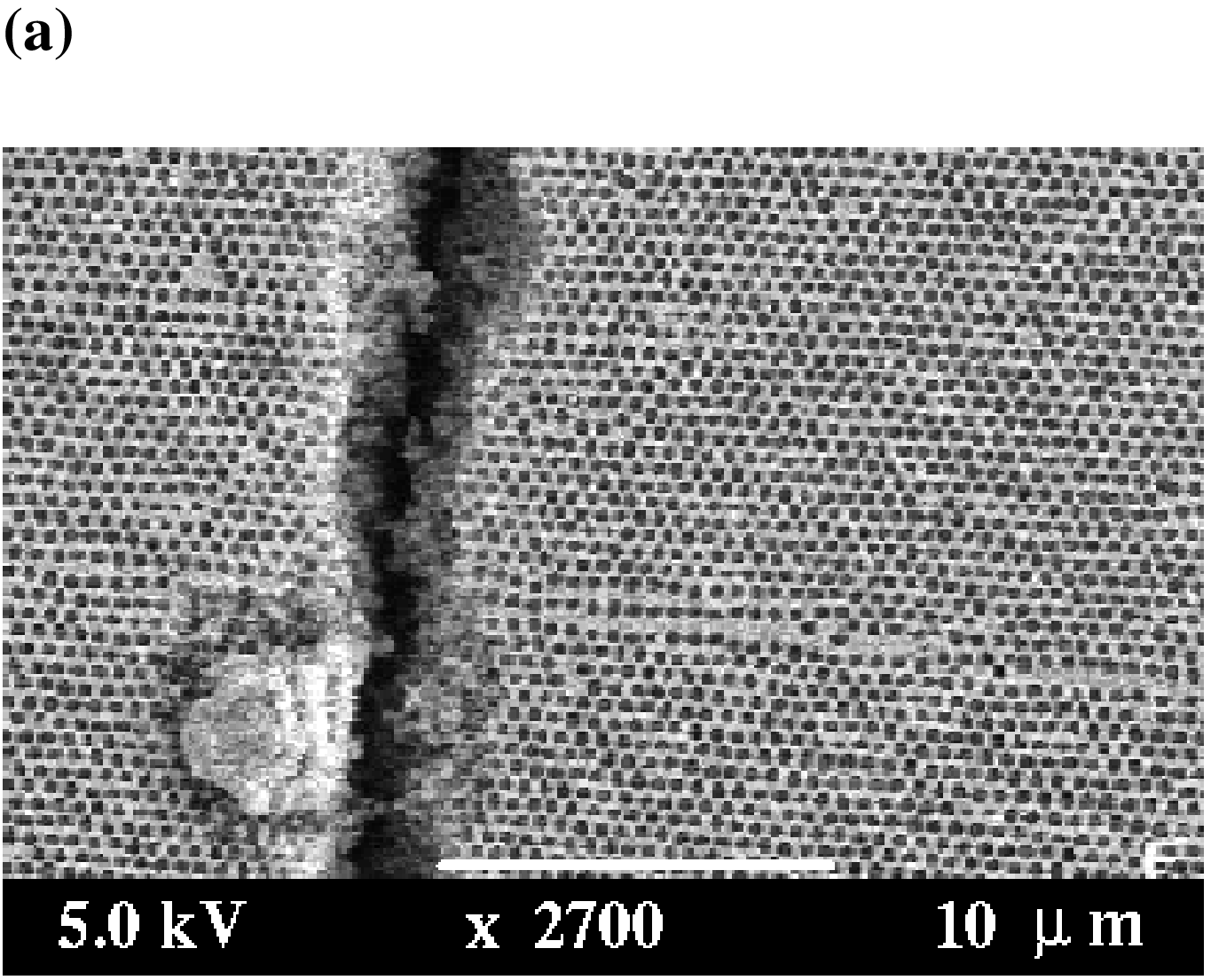}
\includegraphics[width=8cm]{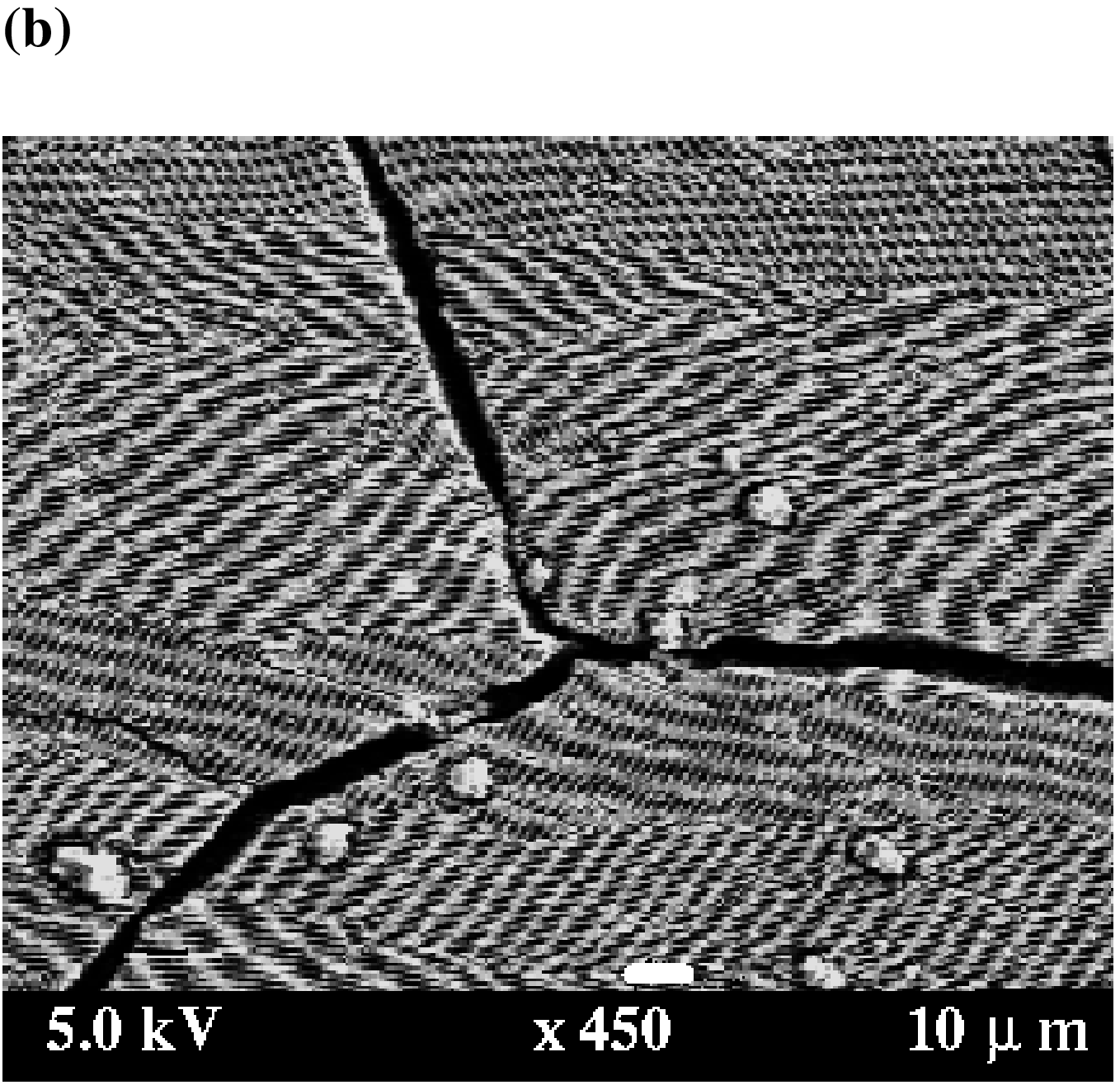}
\includegraphics[width=8cm]{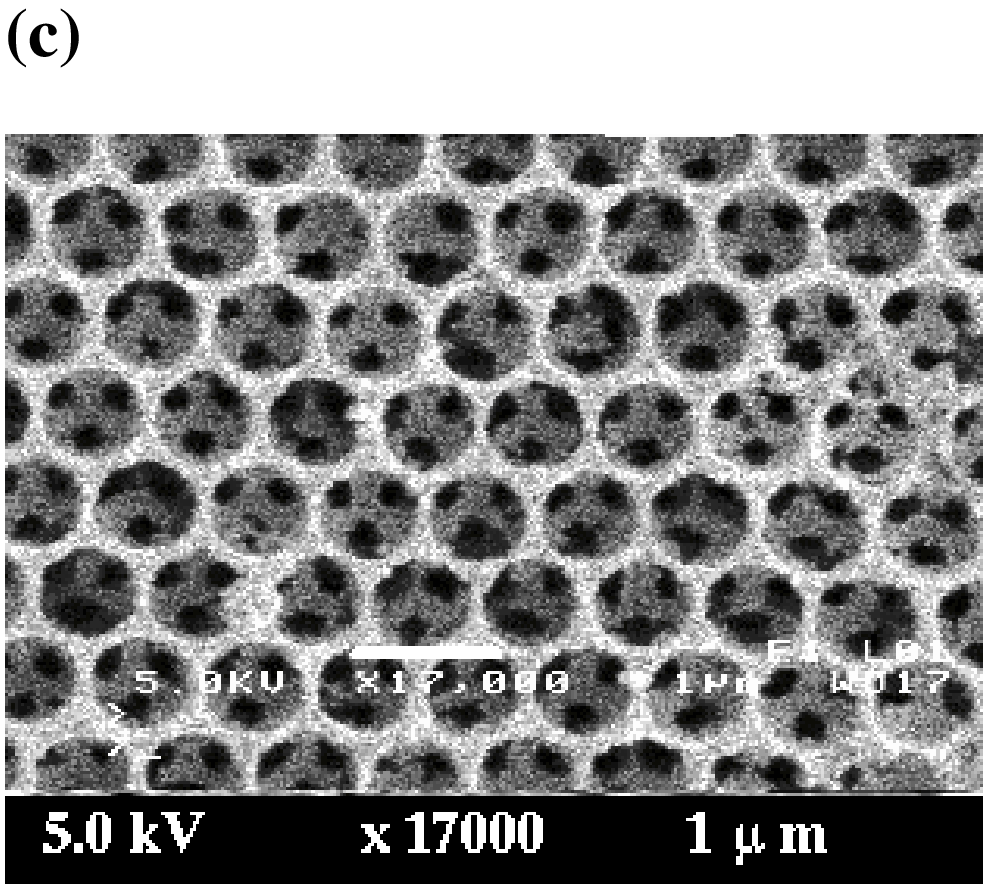}
\caption{(a) SEM micrograph of a PC made using 556 nm polystyrene spheres 
showing excellent long range order. The bar denotes 10 $\mu$.
(b) Low magnification (x450) SEM image showing the Moire pattern 
illustrating the domain orientation
and strain within the domain. The bar represents 10 $\mu$.
(c) A high magnification (x17,000) SEM image showing the close-packing 
of the air spheres. The PC was fabricated from 770 nm polystyrene spheres.
The bar represents 1 $\mu$.
The dark regions inside each hollow region correspond to the position of the spheres
in the next layer.}
\end{figure}

\begin{figure} 
\includegraphics[width=7cm]{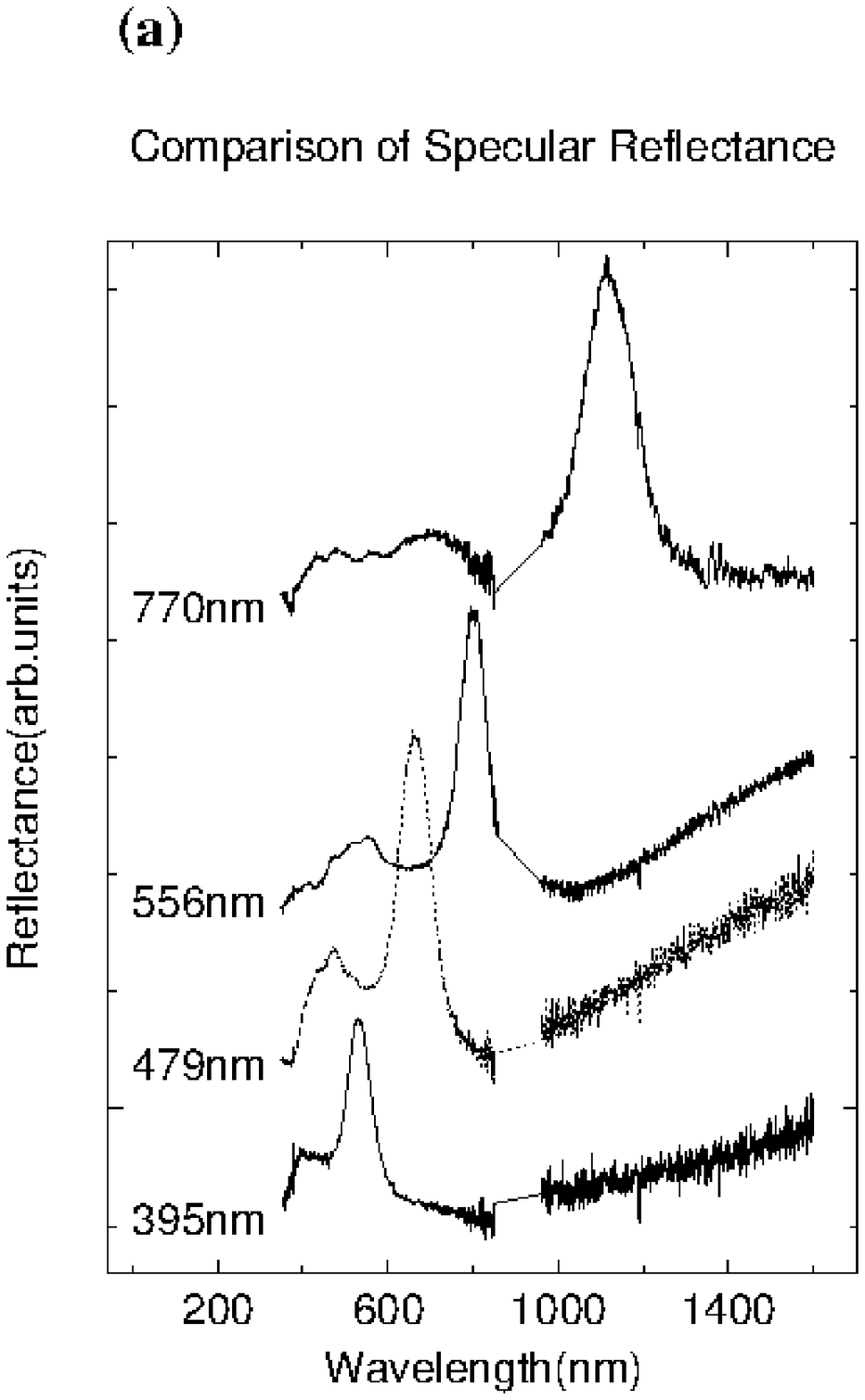}
\includegraphics[width=7cm]{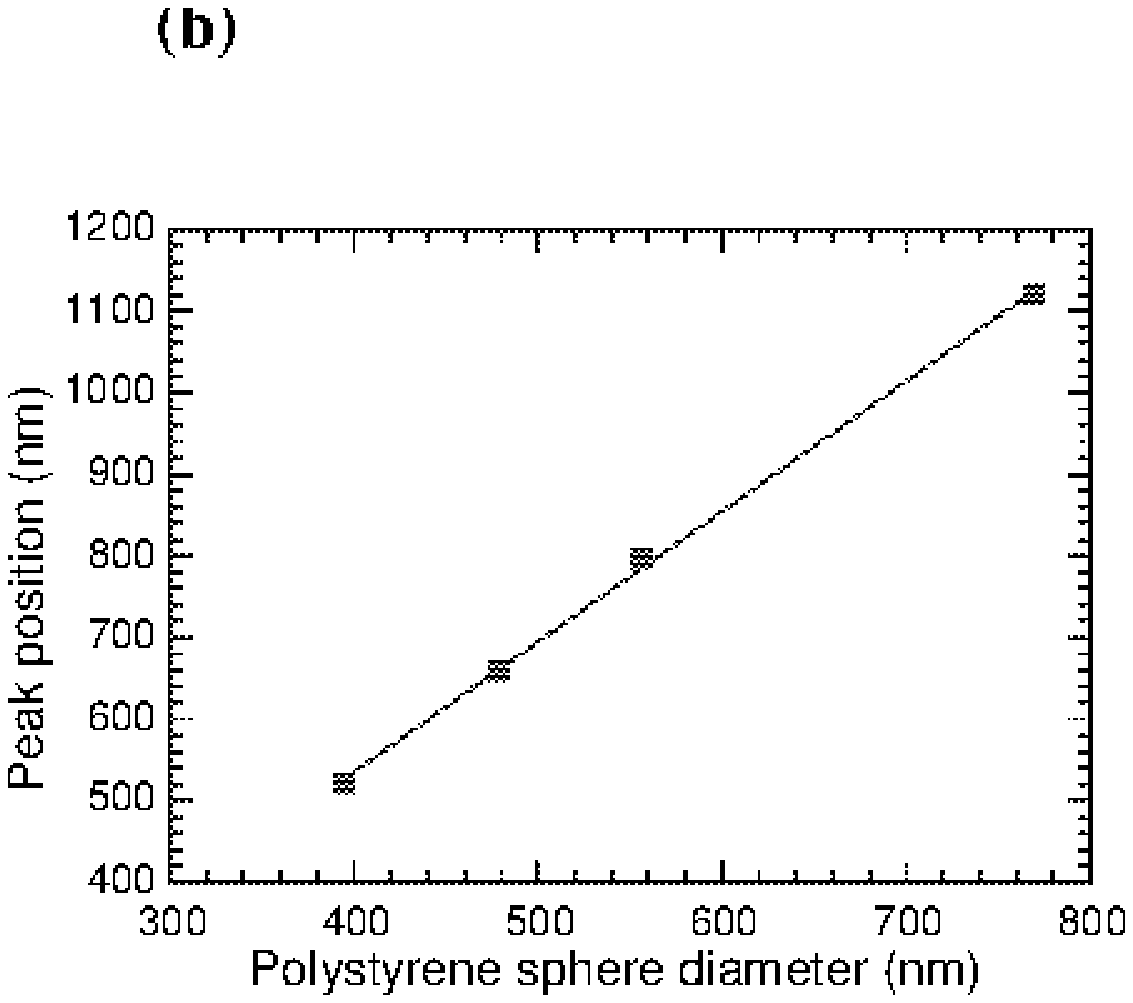}
\caption{(a) Specular reflectivity measurements from photonic crystals, 
fabricated from spheres of varying diameter listed in the legend. The systematic shift in the reflectivity of the first peak is a true optical signature of the photonic crystals.
(b) Wavelength of the primary reflectivity peak as a function of the diameter of the polystyrene sphere templates. The straight line is a linear fit to the peak wavelengths. Very good scaling is observed with sphere diameter.}
\end{figure}

\end{document}